\documentstyle[aps,prl,epsfig,multicol,times]{revtex}
\makeatletter
\newlength\abovecaptionskip \newlength\belowcaptionskip
\def\@makecaption#1#2{%
 \vskip\abovecaptionskip \sbox\@tempboxa{#1: #2}%
 \ifdim \wd\@tempboxa >\hsize #1: #2\par \else \global \@minipagefalse
 \hb@xt@\hsize{\hfil\box\@tempboxa\hfil}%
 \fi \vskip\belowcaptionskip} \makeatother
\begin{document}
\title{ Conductance peak motion due to a magnetic field in weakly 
coupled chaotic quantum dots} 
\author{Ilya L .Kurland$^{1}$, Richard Berkovits$^{1,2,3}$ and 
Boris L. Altshuler $^{1,2}$}
\address{$^{1}$ Physics Department,
Princeton University, Princeton, NJ 08544}
\address{$^{2}$ NEC Research Institute, 4 Independence Way,
Princeton, NJ 08540}
\address{$^{3}$ Minerva Center and Department of Physics,
Bar-Ilan University, Ramat-Gan 52900, Israel}
\date{\today, draft 1.0}
\draft 
\maketitle
\begin{multicols}{2}[%
\begin{abstract} 
We study the influence of moderate exchange interactions of
electrons on the behavior of the peaks in the
conductance of single electron transistors. 
We numerically reproduce recently observed features of both the peak
positions and the peak heights magnetic field dependence.
These features
unambiguously identify the total spin $S$ of each ground state.
We evaluate the probability of each $S$
(combination of spontaneous and induced
magnetization) as a function of the exchange strength, $J$, and external
magnetic field , $B$. The expressions involve only $J$ and 
$g$-factor as adjustable parameters. Moreover, in a surprisingly broad
parameter range these probabilities are determined by certain
linear combinations of $J$ and $B$. 
\end{abstract}
]

In the Pauli picture, electrons populate the orbital states 
of a chaotic quantum dot
or metallic grain
in a sequence of spin up - spin down electrons.
This means that the total spin of the system 
can be either zero (when the number of electrons $N$
is even) or one half (at odd $N$). 
It is well known, \cite{dotreview} 
that measuring the electron transport through
a quantum dot one effectively studies its
ground state (GS) properties. 
One can measure, e.g., the ohmic
conductance by weakly connecting the dot with 
source and drain electrodes. 
This conductance can be studied 
as a function of the voltage, $V_g$,
applied between the dot and 
a third electrode - gate.
The energy of the system  
at a given number of electrons, $E_N$, 
is determined by the gate voltage, $V_g$.
Each time when $E_{N}(V_g)$ 
coincides with  $E_{N+1}(V_g)$
the conductance increases dramatically.
Such a coincidence determines the position, $V_{g}^{(N)}$
of $N$-th peak in the conductance, which is
proportional to $\mu_N$, determined as
$\mu_N = E_{N}(0)-E_{N-1}(0)$.
Accordingly, the distance between the peaks
is $\propto \mu_N - \mu_{N-1}$.
The height of the conductance peak 
is determined by 
the GS wave functions.
Thus, 
studies of the conductance as function of $V_g$ 
allow to deduce properties 
of the {\it manyparticle} GS.

\begin{figure}\centering
\vskip -1.3truecm 
\epsfxsize7cm\epsfbox{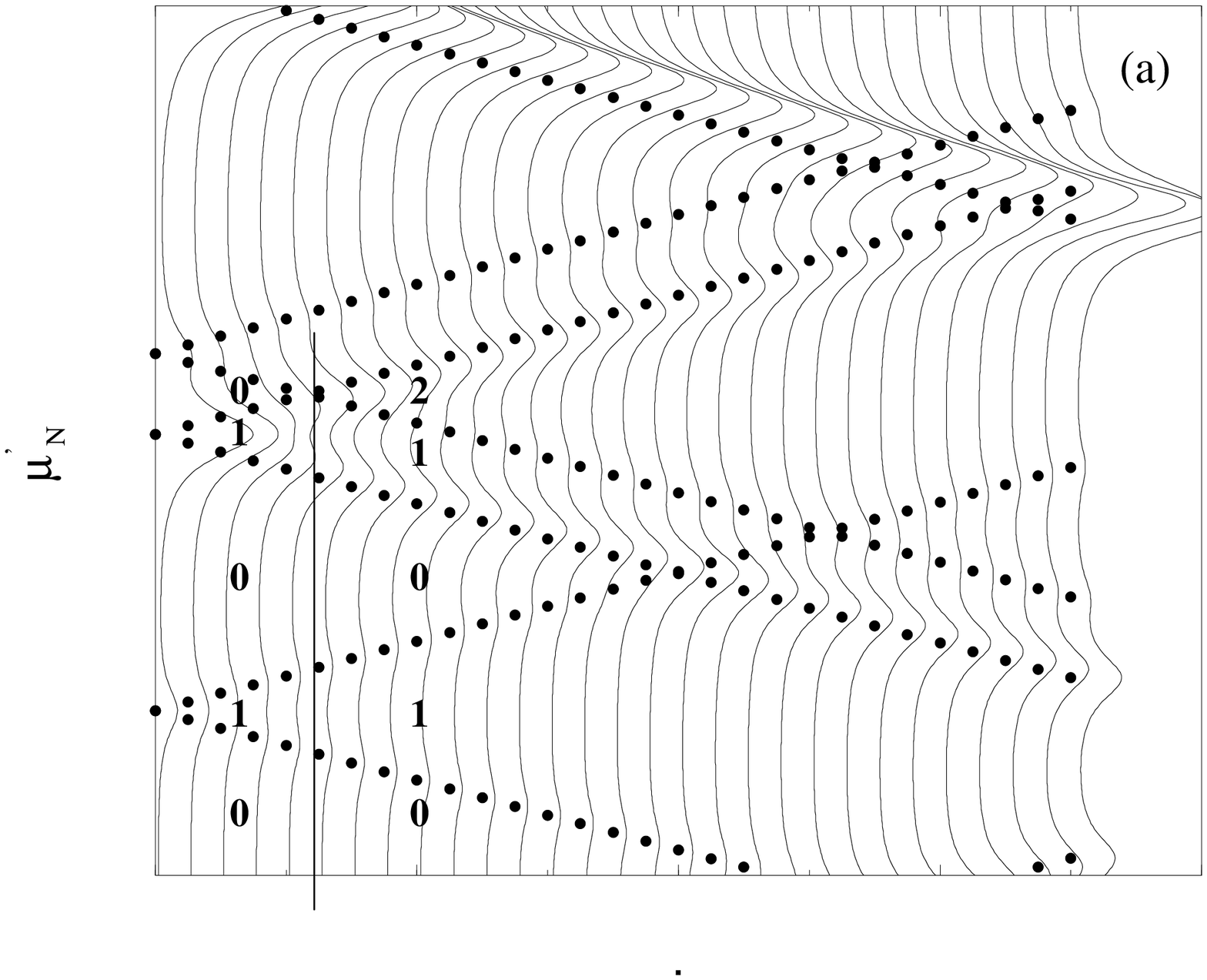}
\vskip -2.2truecm
\epsfxsize7cm\epsfbox{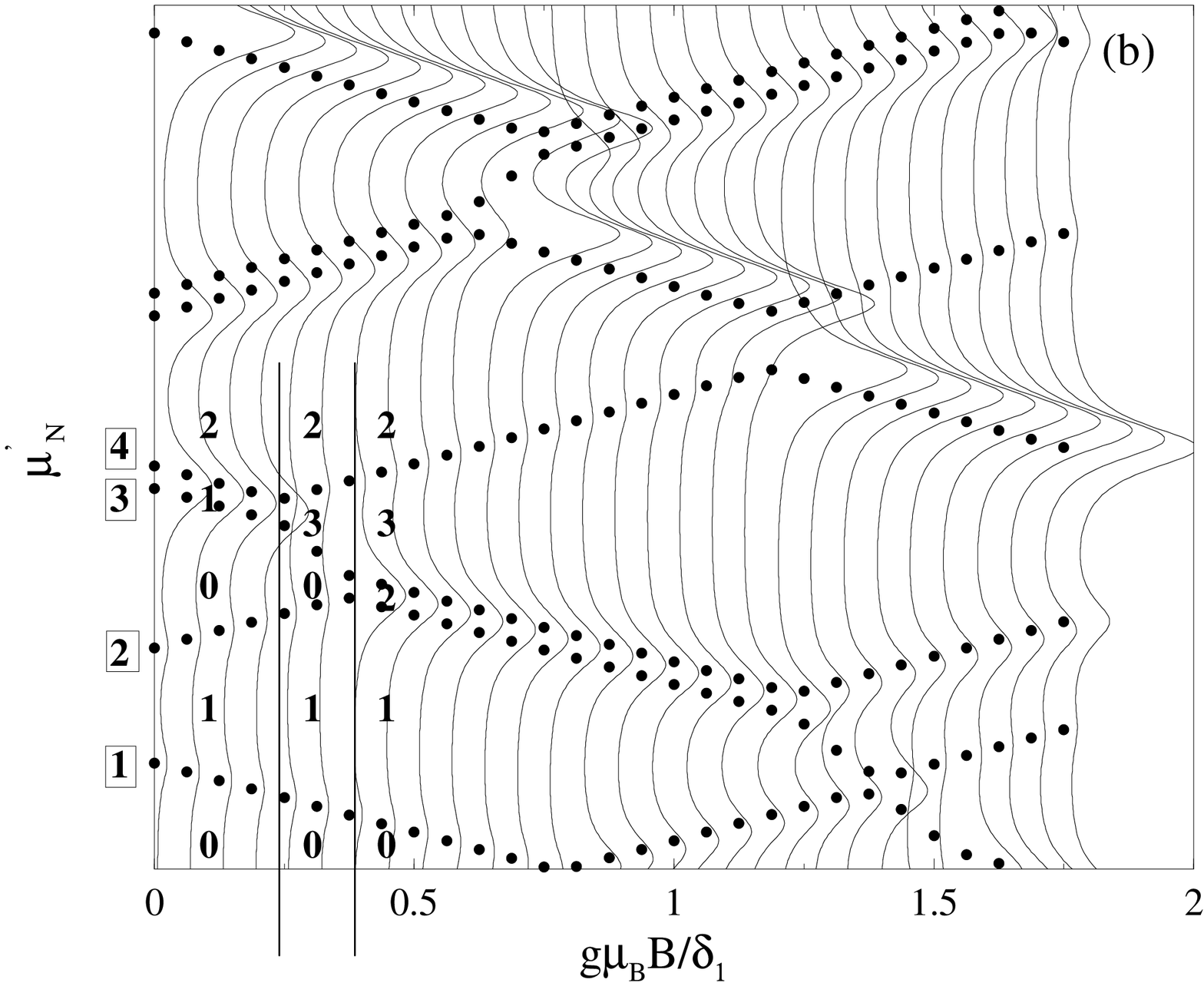}
\vskip -0.3truecm
\parbox{8.5cm}{\caption[]{\small 
Typical conductance dependence on the gate voltage
and magnetic field, $B$, with and without exchange interaction:
(a) $J=0$, 
(b) $J=0.1 \delta_1$. 
$\mu'_{N}$ is determined as 
$\mu'_{N}=\mu_{N}-e^2/C$ and
taken in arbitrary units. 
The vertical curves indicate the conductance while 
the circles indicate the peak maxima 
positions. 
The behavior of specific peaks (indicated by $[1]-[4]$) in (b) and the 
different regions of GS spin 
(depicted in the plot in units of $1/2$, i.e., $1$ corresponds to $S=1/2$, and
separated for clarity by vertical lines)
is discussed in the
text. }
\label{fig1}}\end{figure}

Recently, the behavior of the conductance peaks 
in metallic dots \cite{ralph95}, carbon nanotubes \cite{cobden98},
small chaotic GaAs \cite{marcus00} and Si \cite{rokhinson00} quantum dots
was monitored as a function 
of $B$.
In the following discussion we neglect the magnetic field
coupling to the orbital degrees of freedom 
(For a 2D quantum dot one is allowed to do it when the 
magnetic field is inplane. 
For an ultrasmall dot this is 
a good approximation for a
field in any direction, provided that
it is small enough). 
In other words, we expect the field
to manifest itself only through the
Zeeman splitting - the field shifts 
the GS energy by $g \mu_B S B$, 
where $\mu_B$ is the Bohr magneton 
and $S$  denotes
the GS total spin.
For systems studied in 
Refs. \cite{marcus00,rokhinson00}
the Zeeman splitting for $S=1/2$
becomes comparable to the mean 
single electron level spacing
\begin{equation} 
\delta_1 = <s_i>; \ \ \  s_i = \varepsilon_{i} - \varepsilon_{i+1},
\label{spacing}
\end{equation}
(here $\varepsilon_{i}$ denotes
the orbital energy of the one-electron 
orbital state $i$, and $<\ldots>$ stands for
the averaging.)
at $B \sim 1$ tesla for the GaAs \cite{marcus00} dot and  $B \sim 10$ tesla 
for the Si dot \cite{rokhinson00}. 
Thus, up to a few Tesla $g \mu_b B \ll \delta_1$,
i.e., in the Pauli picture the magnetic field 
only seldomly exceeds $s_i$ 
resulting in changes of the GS spin.
As a result one should
expect the following conductance peak behavior
as function of the magnetic field
(see Fig. \ref{fig1}a):
(i) The peak positions ($\mu_N$) 
vary with $B$ as consecutive pairs
creating an altering pattern 
of downward and upward moving peaks. 
The trajectories of the peaks are
straight lines with the same slope of $g \mu_b / 2$. 
(For the peak spacings ($\Delta_N$)
the Pauli picture predicts
a similar pattern with slopes of $g \mu_b$). 
Spin-orbit interactions may lead to fluctuations in the $g$
factor resulting in fluctuations in the slope magnitudes
\cite{matveev00,brouwer00}, which will not change the pattern
of downward and upward moving peaks.
(ii) The peak heights are the same 
for each pair and change between neighboring pairs.
(iii) Once $g \mu_b B$ exceeds
a particular spacing $s_i$,
the peaks should cross.
The peak height 
is expected to
switch at each crossing.

Experimental results confirm this picture for
metallic dots and carbon nanotubes\cite{ralph95,cobden98}, but
contradict it for the semiconducting chaotic dots 
\cite{marcus00,rokhinson00}.
In Fig. \ref{fig1}b
we show the behavior of the conductance peaks 
as function of the magnetic field 
that follows from a 
model discussed later. 
Fig. \ref{fig1}b reproduces
features of the semiconducting chaotic dot
experiments (see e.g., Figs. 2 and 3
in Ref. \cite{rokhinson00}) 
much better
than Fig. \ref{fig1}a.
Peaks move as function of the magnetic field
in bunches of two \cite{marcus00,rokhinson00} 
(peaks [3] and [4] in Fig. \ref{fig1}b)
in the same direction and with the same slope.
For these peaks the 
height does not come in pairs. 
Not every change
in the direction of the motion of a
peak looks like a
crossing. 
For example, some of these
changes are more abrupt 
- a particular example is peak [3].
Thus, a description of the system beyond the Pauli picture
is needed.

This letter aims to explain
the puzzling features of the magnetic field 
dependence of the conductance peaks positions 
and heights 
by spontaneous spin polarization 
of the electrons in the dot.  
The possibility of 
spontaneous magnetization has 
been in the center of much recent research 
\cite
{andreev98,berkovits98,brouwer99,eisenberg99,baranger00,jacquod00,kurland00}.
It has been previously suggested as an explanation \cite{berkovits98} 
to the 
puzzling absence of a 
bimodal distribution in the conductance peak 
spacings \cite{exp_spacing}.
Such a magnetization was also
predicted to lead to
kinks in the parametric motion of the peaks, 
(e.g.,
due to orbital effects of a perpendicular magnetic
field) \cite{baranger00}.
Authors of a recent paper
based on a random interaction model 
\cite{jacquod00} note some deviations 
of the peak motion from the Pauli picture 
due to spontaneous magnetization. 

The magnetization results from the 
exchange interaction between the electrons. 
In the presence of interaction 
the Pauli picture
no longer holds. 
Indeed, it might be advantageous to 
lose in kinetic energy by placing electrons
at higher 
orbitals in order to gain in
the exchange energy. 

A symmetric dot is spontaneously polarized
every time there is a partially filled 
degenerate orbital - this
is the familiar Hund's rule in atomic physics. 
Accordingly, magnetization will occur 
in a chaotic dot
once the orbitals are close enough to each other.
Here we demonstrate that
the exchange interaction explains 
the deviations from the
Pauli picture listed above and also make
definite predictions on the statistics
of the peak motions which 
can be checked experimentally.

We use the
following Hamiltonian 
to describe the electrons in a 
chaotic dot \cite{kurland00}:
\begin{equation}
H= \sum_{i} \varepsilon_{i} n_{i} +  {{e^2 N^2} \over {2 C}} -
J S(S+1) + g \mu_b S B.
\label{hamil}
\end{equation}
The first term in Eq.~(\ref{hamil})
is the one-partical Hamiltonian, where
$\varepsilon_{i}$ and $n_{i}$  are the
energy of the i-th  orbital 
and its occupation number.
We assume that the energies $\varepsilon_{i}$ 
are characterized by the Random Matrix spectral
statistics. Here we consider both orthogonal
{\bf (GOE)} and unitary {\bf (GUE)} Dyson ensembles.
The second term represents the charging energy. 
We ignore fluctuations in
the capacitance $C$ \cite{exp_spacing}
which are not important for the magnetic behavior.
It was demonstrated in Ref.~\cite{kurland00}
for the most general form of 
the electron - electron interaction that
the sample to sample and 
level to level fluctuations of
the exchange interaction are negligible 
for systems with high Thouless conductance.
This is the reason why the exchange 
energy can be presented in a simple form 
of the third term of Eq.~(\ref{hamil})
with some constant $J$ that does not fluctuate.
(See Ref.~\cite{kurland00} for the details.).
Note that Refs.~\cite{brouwer99},\cite{jacquod00}
introduce different exchange terms. 

We calculated
the conductance using the energies $E_N$ and 
wave functions obtained numerically 
for GOE and GUE random matrix realizations.
An example of the
peak positions and heights 
evolution as functions 
of the magnetic field for a particular
GOE realization is presented  in Fig \ref{fig1} for 
(a) $J=0$ and (b) $J=0.1 \delta_1$.
For both cases we present $\mu_N'=\mu_N - e^2/C$. 

It is not surprising that in Fig \ref{fig1}a one 
can see all the previously described features of the
Pauli behavior.
On the other hand, once even a
weak exchange interaction is included 
the behavior changes qualitatively. 
For example, even in the vicinity
of $B=0$, around a $S=1$ state 
(peaks [3] and [4] in Fig. \ref{fig1}b) the usual
Pauli pattern is disrupted. 
Instead of an up and down moving pairs 
of identical peak height 
we see two down moving and then 
two up moving peaks with alternating height. 
The reason for this behavior is 
that first two consecutive orbitals 
are first filled with down spin electrons
and only later they acquire  up electrons. 
Generally the enhancement 
of the spin of the dot by $S$
will be accompanied by a $2S$ bunch 
of alternating height peaks moving with the same slope.
If the peak spacing $\Delta_N$ is plotted, 
two sets of $2S-1$ flat curves sandwiching 
a sloped one will appear.
Changes to the crossing patterns 
should be also attributed to the
spontaneous magnetization. 
For example, such a case 
occurs when the spin of the 
GS between peaks $[3]$ and $[4]$
(indicated in Fig. \ref{fig1}b) 
switches from $S=1/2$ to $S=3/2$. 
Peaks $[3]$ and $[4]$ abruptly change their slopes 
and amplitudes 
(actually the amplitude
vanishes due to spin blockade \cite{weinman95}).
The amplitude of peak $[4]$ 
is equal to the amplitude 
of peaks $[1]$ and $[2]$ since
the orbital, which occupation was documented
by appearance of the $[1]$ and $[2]$ peaks,
becomes once again 
available for tunneling through
the combined  effect of
spontaneous and field-induced magnetization.
Another change of slope and amplitude is seen
once the GS between peaks $[2]$ and $[3]$
switches from $S=0$ to $S=1$.
Thus, the exchange interaction 
can explain some of the qualitative
behaviors seen in the experiment.

\begin{figure}\centering 
\vskip -1.2truecm
\epsfxsize6.6cm\epsfbox{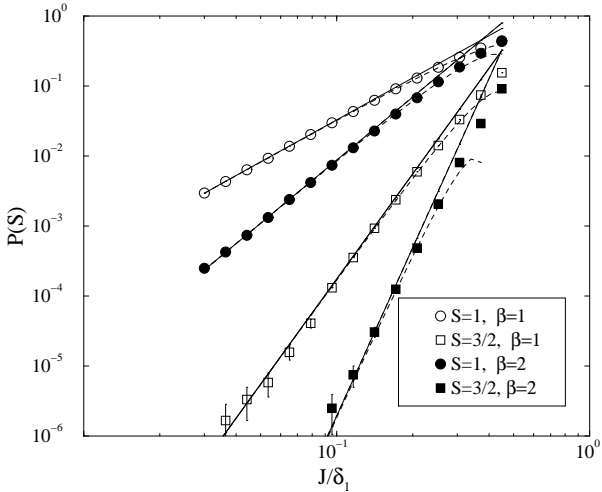}
\vskip -0.3truecm
\parbox{8.5cm}{\caption[]{\small 
Probabilities $P_{J}(S)$ of 
$S=1$ and $S=3/2$ at $B=0$ 
as functions of the exchange 
strength $J$ for GOE
($\beta=1$) and 
GUE ($\beta=2$).
The symbols represent 
numerical results,   
the solid curves correspond to the first order terms in
Eq. (\ref{px}) while the dashed lines 
involve the higher order 
corrections. }
\label{fig2}}\end{figure}

This model can also give quantitative predictions 
regarding the peak position movements.
As illustrated above, the appearance of 
higher spin clearly manifests itself 
in these trajectories.
Using Eq.(\ref{hamil}) 
one can predict
the frequency of spontaneous 
magnetization appearances. 
For weak exchange, 
$J \ll \delta_1$, the probability that 
the dot GS 
has a spin $S$
is determined by the probability of finding $2S$ 
orbitals so close to each other
that the 
gain in exchange energy (due to the polarization)
overwhelms the loss in the kinetic energy (due to
single rather than double occupations of the orbitals).

Using methods described in details
in\cite{mehta91} one can show that 
a sequence of $2S$ levels with a total energy
smaller than $J S(S+1)$ which is necessary for
the spin to reach the value $S$
at $J \ll \delta_1 $
takes place with the probability:
\begin{equation}
P_{J}(S) = C_S^{\beta}\left({{J}\over{\delta_1}} \right)^{(\beta S + 1)(2S-1)}
\left(1-K_S^{\beta}{{J^2}\over{\delta_{1}^2}}\right).
\label{px}
\end{equation}
Here $\lfloor R \rfloor$ denotes the integer part of 
a real number $R$ and, as usual,  $\beta =1(2)$ 
corresponds to GOE (GUE).
In Eq. (\ref{px}) we took into account
the main asymptotics of $P_{J}(S)$ at $J\to 0$
and the first order correction in $J^2/\delta_{1}^2$.
The coefficients $C_S^{\beta}$ and $K_S^{\beta}$ 
depend on both $\beta $ and $S$. 
Their values for $S = 1, 3/2$ 
are presented in table I.

According to  Eq.(\ref{px})
the probability of a GS with
a certain value of $S$
depends on $\beta $ and goes down 
rapidly as $S$ increases. In Fig. \ref{fig2} we
present the probabilities of 
$S=1$ and $S=3/2$ realizations
as functions of the exchange 
strength, $J$, for the GOE ($\beta=1$) and GUE ($\beta=2$).
The probabilities are
obtained from a direct numerical 
solution of Eq. (\ref{hamil})
(symbols) as well as 
from the analytical estimation, Eq. (\ref{px}).
One can see that for $J \leq 0.3$ both results coincide. 

As already mentioned, one can unambiguouslydetermine
the spin 
of each GS
from the trajectories of the conductance peaks.
This fact creates an opportunity to
extract $J$ from 
the experimentally determined
probabilities of different values 
of $S$. 
Note, that it is only one 
dimensionless parameter $J/\delta_{1}$ that
controls the probabilities $P_{J}(S)$ of all
spins at both possible values of $\beta $.

\begin{figure}\centering 
\vskip -1.2truecm
\epsfxsize6.6cm\epsfbox{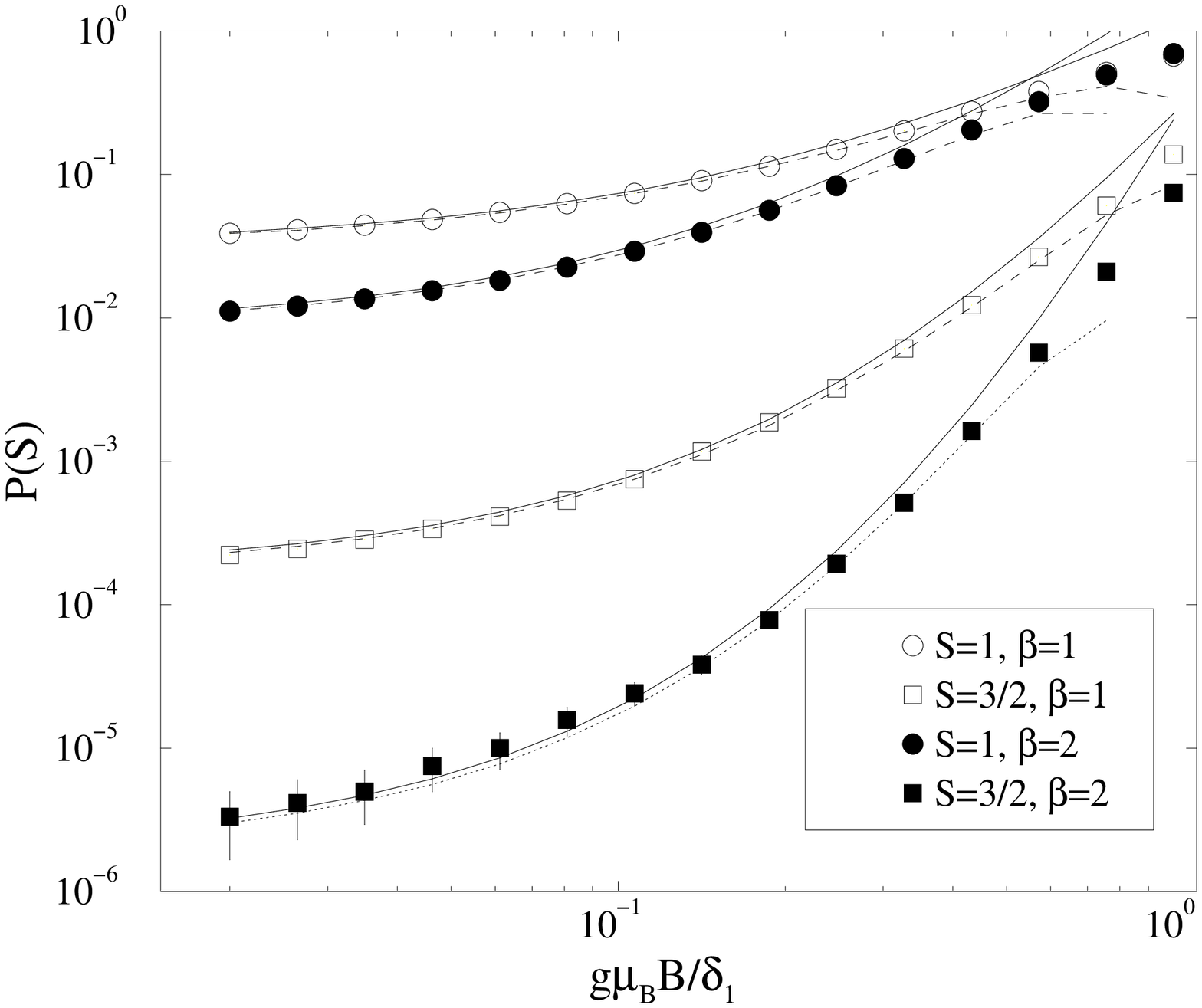}
\vskip -1truecm
\epsfxsize6.6cm\epsfbox{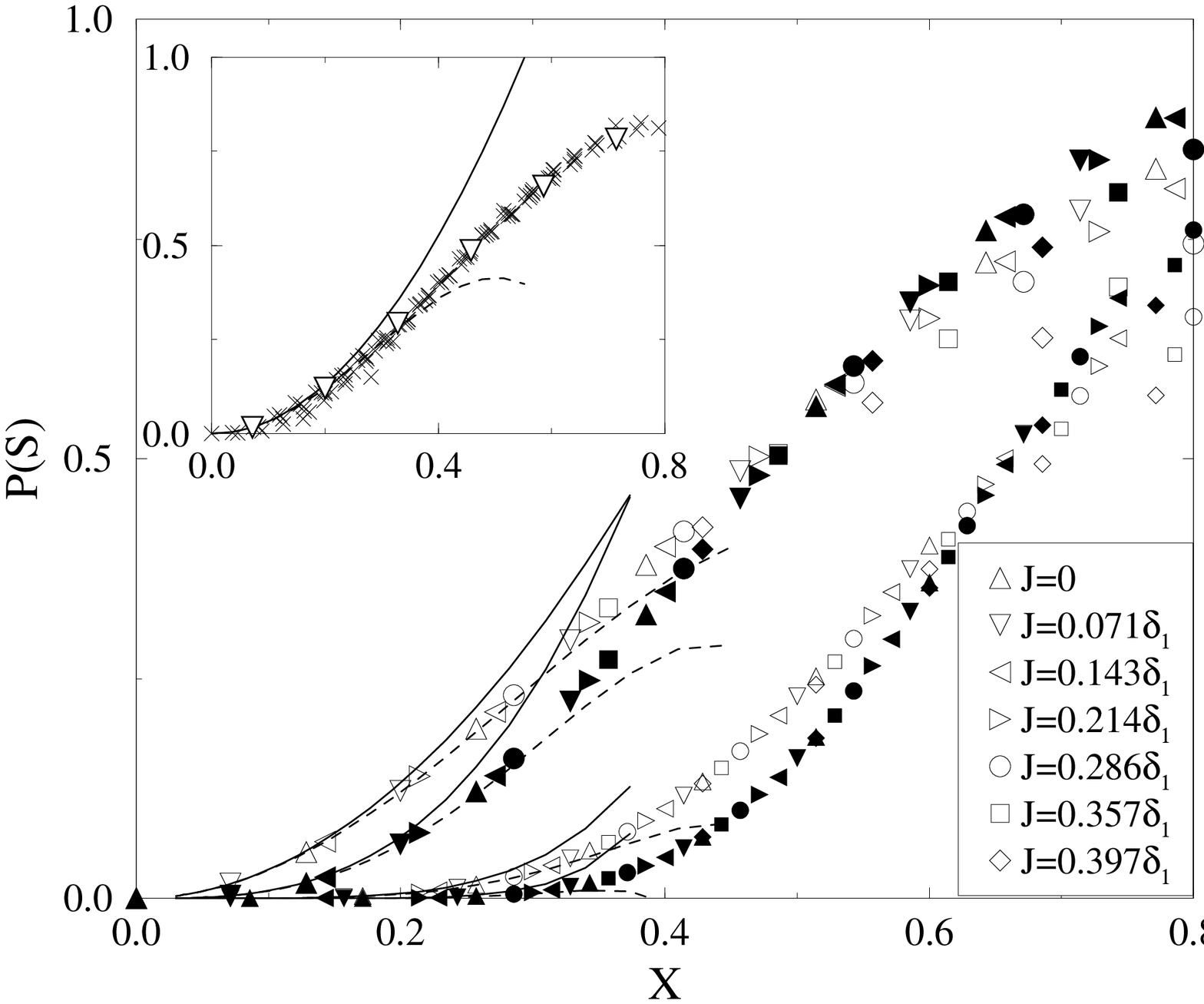}
\vskip -0.3truecm
\parbox{8.5cm}{\caption[]{\small 
(a) Probabilities of 
$S=1,3/2$ for $J=0.1\delta_1$ 
as functions of the
magnetic field $B$. 
The symbols represent 
numerical results,
while the curves (solid - first order, 
dashed - second order)
represent Eq. (\ref{ph}).
(b) The same 
probabilities 
of measuring a spin value $S$ 
for different values of $B$ and $J$ 
scaled as functions of
the parameter $X$, (Eq. (\ref{xx})).
Hollow symbols represent the GOE and filled symbols stand for the GUE.  
Large (small) symbols correspond to $S=1$ ($S=3/2$).
One can clearly see that 
the scaling form holds much beyond the first couple of
terms estimated here.
Inset: probability of $S=1$, in a $\beta=1$ Hubbard model. The Hubbard model
was solve using exact diagonalization (similar to Ref. \cite{berkovits98})
for different lattices and fillings
($4\times3$, $4\times4$, $4\times5$ with $4,6,8$ electrons ),
and different values of on-site interactions, $U=0-2t$,
and magnetic fields. 
The scaling parameter $X$ was calculated using $J(U)$ deduced from the 
spin dependence of the averaged energy.  
The $\times$ symbols correspond to the Hubbard numerical results,
the $\bigtriangledown$ are the same as in (b).
A full description of the Hubbard model results will be given elsewhere. 
}
\label{fig3}}\end{figure}

Magnetic field dependence of the probabilities
of a given spin realization, $P_{J,B}(S)$, can
be calculated in a similar way. It turns out
that as long as the following linear combination
of $J$ and $B$,
\begin{equation}
X={{J}\over{\delta_1}} + {{g \mu_B B} \over 
{\lfloor S+{{3}\over{2}} \rfloor  \delta_1}},
\label{xx}
\end{equation} 
remains small, the probabilities behave as
\begin{equation}
P_{J,B}(S) = C^{\beta}_S
X^{(\beta S + 1)(2S-1)}(1-K_S^{\beta}X^2),
\label{ph}
\end{equation}
In Fig. \ref{fig3}a the 
probabilities $P_{J,B}(S)$ for $S=1,3/2$
calculated numerically at $J/\delta_{1}=0.1$
are compared with the asymptotics
of Eq. (\ref{ph}).
The agreement turns out to be good
as long as $g \mu_B B S < 0.4 \delta_1$. 

The form of Eq. (\ref{ph})
and the higher order terms
suggests that exchange interaction and 
Zeeman splitting enter the
probabilities $P_{J,B}(S)$ 
only through the combination $X$.  
As can be seen
from Fig. \ref{fig3}b, a general scaling
\begin{equation} 
P_{J,B}(S) = F(S,X), 
\label{scaling}
\end{equation}

where $F(S,X)$ is some function
of the scaling parameter $X$ and 
the spin $S$, 
holds for values of $X$ which much exceeds
the range of validity of Eq. (\ref{ph}).
Deviations from the scaling law 
Eq. (\ref{scaling}) at higher $X$
are probably due to the fact that 
the probability to observe even higher spins
($S>3/2$) becomes already substantial.
At $J$ close to $\delta_{1}$ one should
use the approach of Ref.\cite{kurland00} to
describe $P_{J,B}(S)$ analytically.
It is important to stress that this scaling will hold for any system
of interacting electrons as long as the Thouless energy is large and there
is no spin orbit scattering. This is demonstrated in the inset of
Fig \ref{fig3}b where results for a Hubbard model are presented.

In conclusion, we have shown that the presence of
even a relatively weak exchange interaction explains many
of the deviations from the Pauli picture seen in recent experiments
\cite{marcus00,rokhinson00}. We calculate
how the frequency at which
different values of spin appear
depends on
weak exchange interaction and 
a moderate magnetic field. 
The strength of the exchange interaction, $J$,
and the effective $g$-factor are the only
adjustable parameters that determine the
probabilities for a GS of the dot
to have any given spin at any given magnetic
field for both GOE and GUE cases.
In particular
we predict that these
probabilities will follow a one parameter scaling 
law over a wide range of magnetic fields and exchange
interaction strengths. To test
these predictions experimentally 
one needs only enough statistics for
the behavior of the conductance peaks. 

The work at Princeton University
was supported by ARO MURI DAAG55-98-1-0270.
We are grateful to I. L. Aleiner, C. M. Marcus 
and L. P. Rokhinson for many useful 
discussions.

\end{multicols}

\vskip -0.8truecm

\begin{table} [t]
{\protect \narrowtext 
\begin{tabular}{|c|c|c|c|c|}
     & $\beta=1$\span\omit &$\beta=2$\span\omit\\ \hline 
     & $S=1$&$S=3/2$&$S=1$&$S=3/2$\\ \hline
$C$  & $\pi^2/3$&$9 \pi^4/ 50$&$8\pi^2/9$&$81\pi^6/400$\\
$K$  & $\pi^2/5$&$18 \pi^2/49$&$8\pi^2/25$&$792\pi^2/1225$\\
\end{tabular}
\caption{\small The factors $C_S^{\beta}$ and $K_S^{\beta}$ 
appearing in Eq. (\ref{px}).}}
\label{tab1}
\end{table}


\vspace{-0.5cm}
\begin{thebibliography} {50}
\vspace{-1.5cm}

\bibitem{dotreview}  M. A. Kastner, Rev. Mod. Phys. {\bf 64}, 849 (1992);
R. C. Ashoori, Nature {\bf 379}, 413 (1996);
P. L. McEuen, Science {\bf 278}, 1729 (1997).

\bibitem{ralph95} D. C. Ralph, C. T. Black and M. Tinkham, 
\prl {\bf 74}, 3241 (1995). 

\bibitem{cobden98} D. H. Cobden et. al., 
\prl {\bf 81}, 681 (1998).

\bibitem{marcus00} C. M. Marcus (preprint)

\bibitem{rokhinson00} L. P. Rokhinson, et. al., cond-mat/0005262.

\bibitem{matveev00} K. A. Matveev, L. I. Glazman and A. I. Larkin, 
cond-mat/0001431

\bibitem{brouwer00} P. W. Brouwer, X. Waintal and B. I. Halperin, 
cond-mat/0002139

\bibitem{andreev98} A. V. Andreev and A. Kamenev, Phys. Rev. Lett. 
{\bf 81}, 3199 (1998).

\bibitem{berkovits98} R. Berkovits, \prl {\bf 81}, 2128 (1998). 

\bibitem{brouwer99} P.W. Brouwer, Y. Oreg and B.I. Halperin, \prb {\bf 60}, R13977 (1999).

\bibitem{eisenberg99} E. Eisenberg and R. Berkovits, \prb {\bf 60}, 15261 (1999).

\bibitem{baranger00} H.U. Baranger, D. Ullmo and L.I. Glazman,  \prb {\bf 61}, R2425 (2000).

\bibitem{jacquod00} P. Jacquod and A. D. Stone, \prl {\bf 84}, 3938 (2000).

\bibitem{kurland00} I. L. Kurland, I. L. Aleiner and B. L. Altshuler, cond-mat/0004205.

\bibitem{exp_spacing} 
U. Sivan, et. al., \prl {\bf 77}, 1123 (1996);
F. Simmel, et. al., 
Europhys. Lett. {\bf 38}, 123 (1997);
S.R. Patel, et. al., \prl {\bf 80}, 4522 (1998); 
F. Simmel, et. al., \prb {\bf 59}, R10441 .

\bibitem{weinman95} D. Weinman, W. H\"ausler, and B. Kramer  \prl {\bf 74}, 984 (1995)

\bibitem{mehta91} M. L. Mehta {\it Random Matrices}, (Academic Press, NY, 1991). 

\end{thebibliography}
\end{document}